%% file: jbagdonaite-manuscript.tex
\def\apjl{Astroph.\ J.\ Lett.\ }

\def\prl{Phys.\  Rev.\ Lett.}

\def\jms{J. Mol.\ Spectrosc.\ }

\def\mnras{Mon.\ Not.\ Roy.\ Astron.\ Soc.\ }

\def\aap{Astron.\ Astrophys.\  }

\def\rmp{Rev. Mod. Phys.}

\documentclass[aps,prl,twocolumn,showpacs,longbibliography]{revtex4-1}
\usepackage{graphics}
\usepackage{verbatim}
\usepackage{hyperref}
\usepackage{url}
\usepackage{amsmath}
\usepackage{natbib}
\usepackage{color}
\def\arcsec{\hbox{$^{\prime\prime}$}}

\begin{document}
\title{A constraint on a varying proton--electron mass ratio 1.5 billion years after~the~Big~Bang}
\author{J. Bagdonaite, W. Ubachs}
\affiliation{Department of Physics and Astronomy, and LaserLaB, VU University, De Boelelaan 1081, 1081 HV Amsterdam, The Netherlands}
\author{M. T. Murphy, J. B. Whitmore}
\affiliation{Centre for Astrophysics and Supercomputing, Swinburne University of Technology, Melbourne, Victoria 3122, Australia}
\date{\today}
\begin{abstract}{A molecular hydrogen absorber at a lookback time of 12.4 billion years, corresponding to 10$\%$ of the age of the universe today, is analyzed to put a constraint on a varying proton--electron mass ratio, $\mu$. A high resolution spectrum of the J1443$+$2724 quasar, which was observed with the Very Large Telescope, is used to create an accurate model of 89 Lyman and Werner band transitions whose relative frequencies are sensitive to $\mu$, yielding a limit on the relative deviation from the current laboratory value of $\Delta\mu/\mu=(-9.5\pm5.4_{\textrm{stat}} \pm 5.3_{\textrm{sys}})\times 10^{-6}$. }
\end{abstract}
\maketitle

The accelerated expansion of the universe is ascribed to an elusive form of gravitational repulsive action referred to as dark energy \cite{Perlmutter2012,*Schmidt2012,*Riess2012}. Whether it is a cosmological constant, inherent to the fabric of space-time, or whether it may be ascribed to some dynamical action in the form of a scalar field $\phi$ \cite{Ratra1988}, is an open issue. In the latter case it has been shown that the interaction of the postulated quintessence fields $\phi$ to matter cannot be ignored, giving rise to a variation of the fundamental coupling constants and a breakdown of the equivalence principle \cite{Carroll1998,Bekenstein2002}. In this context it is particularly interesting to probe possible variations of the fundamental constants in the cosmological epoch of the phase transition, going from a matter-dominated universe to a dark energy-dominated universe, covering redshift ranges $z=0.5 - 5$ \cite{Sandvik2002}. While models of Big Bang nuclear synthesis probe fundamental constants at extremely high redshifts ($z=10^8$) \cite{Berengut2013}, the Oklo phenomenon ($z=0.14$) \cite{Damour1996} and laboratory atomic clock experiments ($z=0$) \cite{Rosenband2008} probe low redshifts. Absorbing galaxies in the line-of-sight of quasars are particularly suitable for investigating the range of medium--high redshifts, for a varying fine-structure constant, $\alpha$, based on metal absorption \cite{Murphy2003} and for a varying proton--electron mass ratio, $\mu=m_{\textrm{p}}/m_{\textrm{e}}$, based on molecular absorption \cite{Jansen2014}. Furthermore, unification scenarios predict that variations of $\alpha$ and $\mu$ are connected, while in most schemes $\mu$ is a more sensitive target for varying constants \cite{Ferreira2014}.

A variation of $\mu$ may be probed through the spectroscopy of molecules such as hydrogen (H$_2$)\cite{Ubachs2007}, ammonia (NH$_3$)\cite{Henkel2009} and methanol (CH$_3$OH)\cite{Bagdonaite2013}. The latter polyatomic molecules are more sensitive to a variation in $\mu$ under the assumption that molecular reduced masses, involving protons and neutrons, scale in a similar manner as $\mu$, which can be probed in a pure form in H$_2$ \cite{Jansen2014}. Moreover, the radio absorption systems, where the polyatomic absorbers can be found, are more rare and are currently only found at the lower redshifts $z=0.68$ \cite{Murphy2008,Kanekar2011} and $z=0.89$ \cite{Bagdonaite2013}. Molecular lines observed at $z = 6.34$~\cite{Riechers2013} are of insufficient spectral quality to constrain $\mu$-variation. Conveniently, the ultraviolet spectrum of the H$_2$ Lyman and Werner bands can be investigated with large ground-based optical telescopes for absorbers at redshifts $z > 2$. The current sample of H$_2$-based measurements covers a redshift interval from 2 to~3~\cite{Weerdenburg2011, Rahmani2013}, with the highest redshift object Q0347$-$383 at $z =3.025$~ \cite{King2008,Wendt2012}. 

\begin{figure*}
\resizebox{0.85\textwidth}{!}{\includegraphics{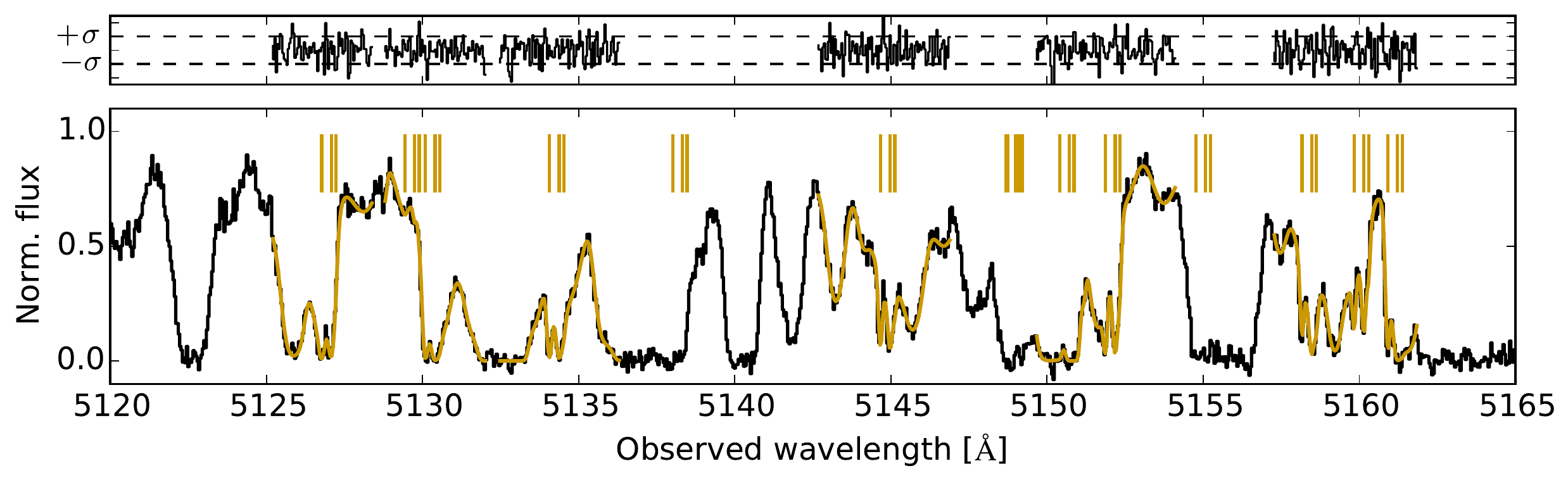}}
\caption{Part of the final, fitted spectrum of J1443$+$2724. The sticks indicate the 3 velocity components of each H$_2$ transition from the rotational levels $J=(0-4)$. The residuals are shown above the spectrum for the regions fitted. The broad features that surround and overlap the fitted H$_2$ transitions are multiple, unrelated H{\sc i} Lyman-series absorption lines arising in other absorbers along the line of sight.}
\label{fig1}
\end{figure*} 

Here we constrain variations in $\mu$ at substantially higher redshift by analyzing an H$_2$ absorber at $z=4.22$ along the sightline towards the background ($z=4.42$) quasar PSS\,J1443$+$2724~\cite{Ledoux2006}. This step to higher redshift is challenging for several reasons. Firstly,  more distant quasars are typically fainter, making initial discovery of the H$_2$ absorption more difficult and requiring longer integration times for a high-quality spectrum with which to constrain $\Delta\mu/\mu$. 
Secondly, absorption lines from neutral hydrogen (H{\sc i}) in the intergalactic medium are more numerous at higher redshifts, complicating analysis of the H$_2$ transitions. 
The Lyman-series transitions from the many unrelated H{\sc i} clouds in the intergalactic medium form a characteristic `forest' of broader spectral features against which the H$_2$ have to be identified and analyzed (see Fig.~\ref{fig1}). However, the `comprehensive fitting' method  of simultaneously treating this H{\sc \i} forest and the H$_2$ absorption, developed previously and documented extensively~\cite{King2008, Malec2010,Weerdenburg2011,Bagdonaite2014}, is employed here to reliably meet this challenge.

A portion of the J1443$+$2724 spectrum is shown in Fig.~\ref{fig1}. To create the spectrum, we make use of an archival dataset obtained in 2004~\cite{Ledoux2006} and a new dataset obtained in 2013 (Program ID 090.A-0304) which comprise, respectively, 7.3 and 16.7 hours (5 and 12 exposures) observation time with the Very Large Telescope/the Ultraviolet and Visual Echelle Spectrograph (VLT/UVES). For weak light sources as the present quasar (visual magnitude $V\sim 19.4$) long integration time is required to reach signal-to-noise ratio (SNR) of $>$30, especially if high spectral resolution is required. Wavelength calibration in the UVES instrument is achieved by observing a reference ThAr lamp. After each night of observations (in 2004) or, preferably, immediately after each quasar exposure (in 2013) a spectrum of the ThAr lamp was recorded using identical settings to those of the science exposure. The ThAr spectrum was used by the UVES data-reduction pipeline to create pixel-vs.-wavelength maps and apply them to the quasar exposures. The wavelength scales of the spectra were converted to vacuum--heliocentric values. Following the same procedures as in~\cite{Malec2010,Weerdenburg2011,Bagdonaite2014}, flux-extracted individual exposures were then resampled, scaled and merged to a final 1D spectrum that extends from 474 to 792 nm, with the H$_2$ transitions detected at \mbox{484--581 nm}. The SNR at shorter wavelengths is entirely dominated by the data from 2004 as the observations in 2013 were affected by stray light from the full moon.

\begin{table}
	\caption{Results of fitting models of increasing complexity to the spectrum. The two velocity features (see Fig. \ref{fig2}) require fitting at least 2 velocity components (VC) but as it can be seen from the goodness-of-fit measure $\chi_{\nu}^2$ (where $\nu$ is the number of degrees of freedom; here $\nu\sim4600$) models with 3, 4, and 5~VC replicate the data better. For each model, a resulting $\Delta\mu/\mu$ value is shown in the last column.}
	\label{results}
    \begin{tabular}{c@{\hspace{5pt}}c@{\hspace{5pt}}c@{\hspace{7pt}}}

	\toprule
    Number of VCs & $\chi_{\nu}^2$   & $\Delta\mu/\mu$ [$\times 10^{-6}$]       \\ 
	\colrule
    2        & 1.396   & $-8.7\pm5.2$\\
    3        & 1.161   & $-6.7\pm5.4$ \\
    4        & 1.139   & $-8.3\pm5.4$  \\
    5        & 1.133   & $-2.9\pm5.2$  \\
	\botrule
    \end{tabular}
\end{table}

\begin{figure}[!ht]
\resizebox{0.45\textwidth}{!}{\includegraphics{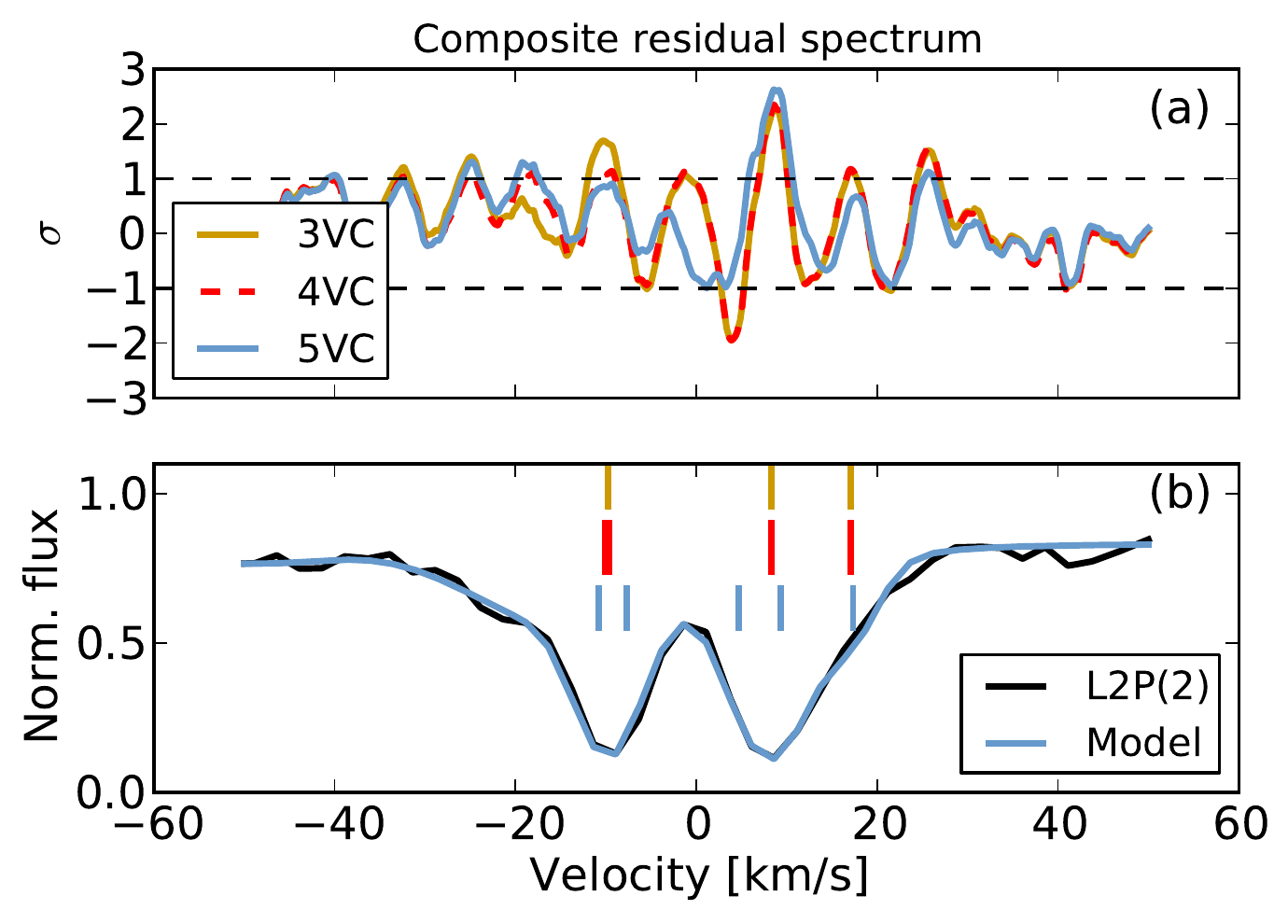}}
\caption{(a) Composite residual spectra of 31 transitions shown for the 3, 4, and 5~VC models. (b) An H$_2$ absorption profile in the J1443$+$2724 spectrum. The sticks indicate the positions of velocity components for the 3, 4, and 5~VC models.}
\label{fig2}
\end{figure}

As shown in Fig. \ref{fig2}, each H$_2$ transition in the J1443$+$272 spectrum has two distinct velocity features. In the comprehensive fitting method employed here, these two features in all H$_2$ transitions are fitted simultaneously (along with the many broader, unrelated H{\sc i} forest features). With this approach, one makes use of the known molecular physics of H$_2$, thereby reducing the number of free parameters in the fit. In particular, transitions from the same ground rotational level $J$ are fitted using a single parameter for column density, $N_J$, while, for the same velocity component (VC), transitions from all $J$ levels are tied in terms of redshift, $z$, and line-width, $b$. The intrinsic intensities of the H$_2$ absorbing transitions are fixed to the oscillator strengths known from the molecular physics database \cite{Malec2010}. In this way, the best fit is achieved by simultaneously combining information from multiple spectral regions. In the case of J1443$+$272, we selected 60 spectral regions containing a total of 89 H$_2$ transitions up to $J=4$, among which 17 are from the Werner band \footnote{See Supplemental Material for the complete list of fitted transitions, the final fitted spectrum, and details on continuum fitting.}. This approach, along with the simultaneous fitting of the H{\sc i} forest lines, is the same as detailed previously (e.g.~\cite{Weerdenburg2011,Bagdonaite2014}). The $\chi^2$ minimization is performed using a Voigt profile fitting program \textsc{vpfit9.5}~\footnote{Developed by R. F. Carswell et al.; available at \mbox{\url{http://www.ast.cam.ac.uk/\string~rfc/vpfit.html}}}. To fit each transition $i$, a Voigt profile is created from a threefold convolution of a Lorentzian profile defined via a damping parameter $\Gamma_i$, a Gaussian profile describing the thermal and turbulent velocities in the gas, and an instrumental profile. For the $i$-th transition of H$_2$ detected at redshift $z_{\rm abs}$, the recorded wavelength is expressed as:
\begin{equation}
\lambda_i = \lambda^{0}_{i}(1+z_{\rm abs})(1+K_i\frac{\Delta\mu}{\mu}),  
\label{eq1}                                                                  
\end{equation}
where $\lambda^{0}_{i}$ is the corresponding laboratory wavelength, and $K_i$ is a coefficient that quantifies the sign and magnitude of its sensitivity to a varying $\mu$ \cite{Ubachs2007}, and where that variation is parametrized by $\Delta\mu/\mu = (\mu_z-\mu_{\textrm{lab}})/\mu_{\textrm{lab}}$.

The two velocity features per transition can contribute towards better precision of $\Delta\mu/\mu$. However, previous studies have shown that an inadequate fit of H$_2$ features can produce a spurious $\Delta\mu/\mu$ value \cite{Murphy2008b}. In particular, underfitted velocity profiles are more prone to such errors. To avoid that, additional VCs are included in the profile to check for a more complex velocity structure than can be appreciated by eye. The additional components can be rejected by \textsc{vpfit} as statistically unjustified or they can remain and improve the model. Table~\ref{results} contains the results of fitting increasingly complex models to the H$_2$ spectrum of J1443$+$272. Fig.~\ref{fig2} displays composite residual spectra of different models. The composite residual spectra are created by combining a number of residuals of individual transitions, which are aligned in velocity/redshift space~\cite{Malec2010}. While according to the reduced $\chi^2$ value, a model with 4 or 5~VCs initially might appear preferable to one with only 3~VCs, fitting the former more complex models was not stable: \textsc{vpfit} rejected the additional VCs in lower-$J$ transitions while retaining them in higher-$J$ transitions, thereby departing from a physically plausible, self-consistent model. Fig.~\ref{fig2} shows that adding a second VC to the left feature of the absorption profile reduces the composite residuals, but that the residuals for the right feature are barely affected when adding a third VC. 

In all the models, the H$_2$ absorption feature on the right requires fitting of at least one very narrow component.
This weak component exhibits an unusually high relative column density for $J=2,3$ levels, and a width of
0.2~km/s, corresponding to a kinetic temperature of 5~K, hence lower than \mbox{$T_{\textrm{CMB}}=14$ K} at \mbox{$z=4.22$.}
While this low $b$ value results from a best-fit model and hence is favored statistically, we performed further testing by imposing $b_{\textrm{min}}$ parameters corresponding to temperatures including those in the expected kinetic regime of \mbox{50--100 K} ($b \sim 0.6-0.9$~km/s).
These tests, performed for 3~VC and 4~VC models demonstrate that resulting values for $\Delta\mu/\mu$ do not critically depend on the narrowness of this velocity feature~\footnote{See Supplemental Material for more information.}. A spectrum of a higher SNR and higher resolution might help in explaining the composite residual excess and the $N, b$ values obtained here. 

Based on the internal consistency of various fitting results, a model with 3~VCs is selected as the fiducial model (see Table~\ref{3VC-model} for fitting results) and is further tested for possible systematics. The 4 and 5~VC fits are used to verify the results of the fiducial model.

\begin{table}
	\caption{Results of fitting a 3~VC model to the data.}
	\label{3VC-model}
    \begin{tabular}{c@{\hspace{5pt}}c@{\hspace{5pt}}l@{\hspace{1pt}}}

	\toprule
    $z_{\textrm{abs}}$ & $b$ [km/s]   & $\log N \pm \sigma_{logN}$ \\ & & from $J=0$ to $J=4$ \\ 
	\colrule
     4.2237292(6)	& 1.74$\pm$0.06 & 17.04$\pm$0.06, 17.40$\pm$0.05, 15.86$\pm$0.09,  \\
     & & 14.81$\pm$0.04, 12.99$\pm$0.84  \\
     4.2240440(11)	& 0.42$\pm$0.18 & 17.49$\pm$0.03, 17.87$\pm$0.02, 17.34$\pm$0.02,  \\
     & & 17.09$\pm$0.03, 13.80$\pm$0.33  \\
     4.2241975(20)	& 0.20$\pm$0.08 & 15.01$\pm$0.33, 16.17$\pm$0.10, 15.77$\pm$0.09,  \\
     & & 15.97$\pm$0.06, 13.67$\pm$0.61  \\
	\botrule
    \end{tabular}
\end{table}

\begin{figure}
\resizebox{0.44\textwidth}{!}{\includegraphics{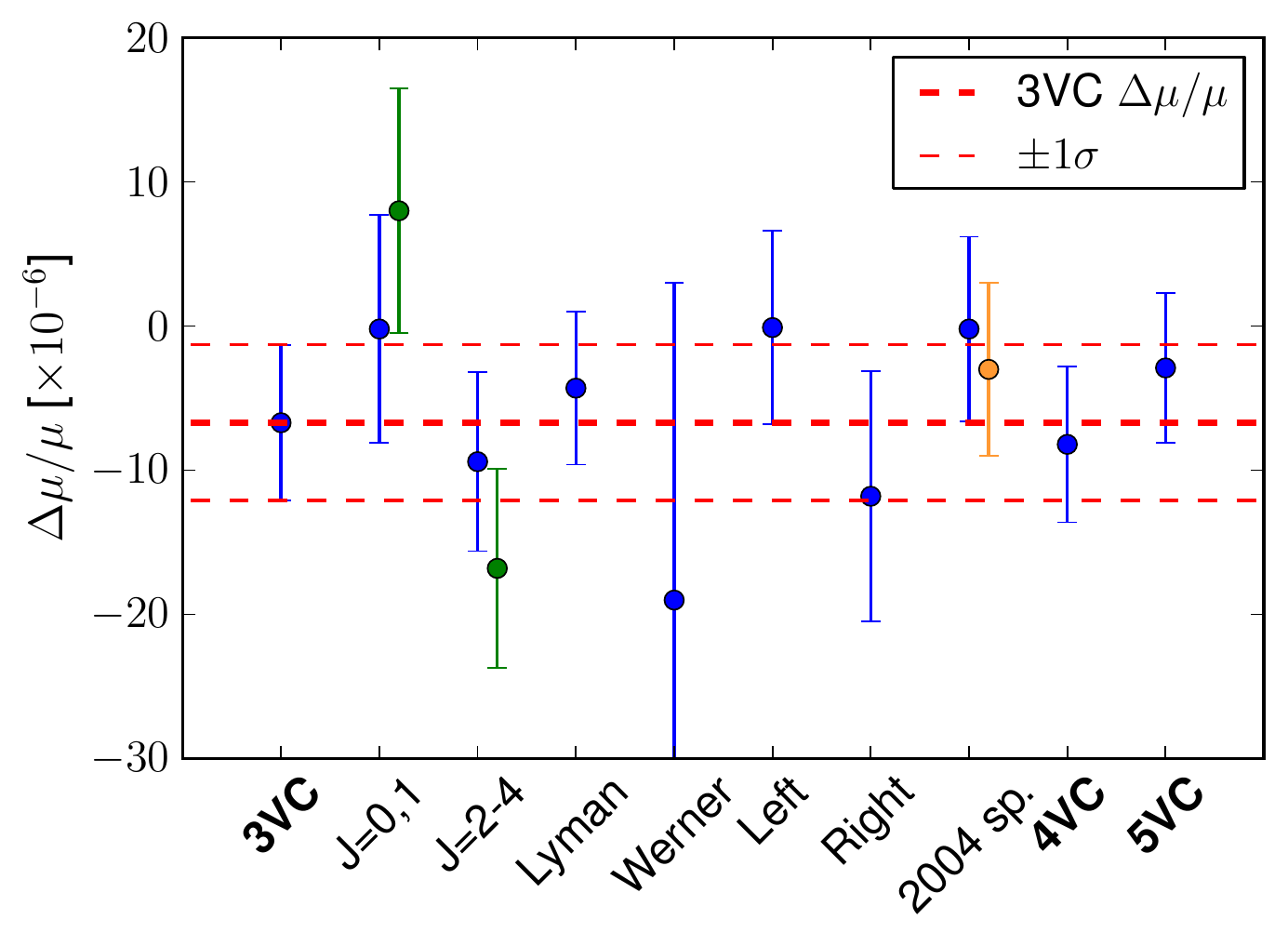}}
\caption{Statistical $\Delta\mu/\mu$ constraints obtained from the fiducial 3~VC model, from various tests performed on it, and from the 4~VC and 5~VC models. The green points refer to fitting low- and high-$J$ transitions with the $z$ and $b$ parameters tied within each group but not between the two groups. The left and the right tests refer to deriving $\Delta\mu/\mu$ constraints from the two velocity features of H$_2$ separately (see Fig.~\ref{fig2}). The orange point was obtained after the spectrum obtained in 2004 has been corrected for long-range wavelength distortions.}
\label{fig3}
\end{figure}

The fiducial 3~VC model delivers $\Delta\mu/\mu = (-6.7\pm 5.4_{\textrm{stat}})\times 10^{-6}$. Fig.~\ref{fig3} shows how this result compares to constraints obtained from fitting only certain selected transitions, velocity components, or parts of the spectrum. Most of the derived constraints are compatible with the fiducial result, demonstrating that the result is robust. 
Nevertheless, a possible discrepancy is that constraints from the low- and high-$J$ transitions seem to move in opposite directions if fitted separately. This tendency is observed in the 4 and 5~VC models as well. If the redshift and width parameters are tied separately for the \mbox{low-$J$} and high-$J$ transitions the resulting $z$ and $b$ parameter values agree well between the two groups. Fitting only the small number of Werner transitions (17), together with their smaller spread in $K_i$ coefficients, results in a large uncertainty in $\Delta\mu/\mu$. Fitting a spectrum that includes only the exposures from 2004 delivers $\Delta\mu/\mu = (-0.2\pm 6.4_{\textrm{stat}})\times 10^{-6}$. This constraint, as well as the constraints derived from the 4 and 5~VC models on a full spectrum, are in agreement with the result from the fiducial model. Fitting a spectrum that includes only 2013 exposures results in $\Delta\mu/\mu = (4.0\pm 12.4_{\textrm{stat}})\times 10^{-6}$. The uncertainty is much larger because the SNR of the data collected in 2013 is affected by additional noise from the full moon.

The accuracy to which $\Delta\mu/\mu$ can be measured strongly relies on accurate wavelength calibration of the quasar spectrum. At any wavelength in the J1443$+$2724 spectrum, the error of the wavelength calibration solution is $\sim70$ m/s which translates to $\Delta\mu/\mu$ of 1$\times$10$^{-6}$ given a spread in $K_i$ coefficients of 0.05~\footnote{Calculated via $\Delta\mu/\mu = 1/\Delta K_i \times \Delta v/c \times 1/\sqrt{n}$, where $n$ is the number of ThAr transitions per one echelle order; typically $n>15$.}. 

An additional source of systematic uncertainty might be caused by wavelength-scale distortions inherent to the instrument. Short-range distortions with a characteristic shape repeating in each diffraction order of the echelle spectrograph were found in UVES~\cite{Whitmore2010}. We estimate its potential to systematically shift $\Delta\mu/\mu$ value by applying a saw-tooth like distortion of $\pm100$ m/s to each order and fitting the recombined spectrum, as in \cite{Malec2010,Bagdonaite2013}. This results in a $\Delta\mu/\mu$ shift as small as $1 \times 10^{-7}$.

Recently, indications were reported that the UVES instrument is also susceptible to long-range wavelength distortions that are able to mimic non-zero $\Delta\mu/\mu$ values at the level of several parts per million \cite{Rahmani2013}. Their origin is not clear but deviations between the light paths of the quasar and ThAr lamp is considered the likely (proximate) cause. The effect of such a miscalibration can be quantified by observing objects that have well-understood spectra, such as asteroids which reflect the solar spectrum and stars with solar-like spectra, known as solar twins \cite{Molaro2008, Rahmani2013, Whitmore2014, Bagdonaite2014}. In the archive of UVES~\footnote{\url{http://archive.eso.org/eso/eso_archive_main.html}}, we found several sun-like stars that were observed in 2004, at a similar time as the J1443$+$2724 quasar~\footnote{See Supplemental Material for the observational details.}. By comparing their spectra to a highly accurate Fourier-Transform spectrum of the sun~\footnote{\url{http://kurucz.harvard.edu/sun/irradiance2005/irradthu.dat}} we found a correction amounting to 44.9$\pm$46.8 m/s/1000\,\AA. 
An extensive study of this particular systematic effect shows that corrections of this size are typical for the considered time period \cite{Whitmore2014}. Applying the aforementioned correction on the wavelength scale of the spectrum obtained in 2004 results in a $\Delta\mu/\mu$ shift of $-2.8 \times 10^{-6}$ if compared to the uncorrected constraint. The uncertainty on this correction translates into a systematic $\Delta\mu/\mu$ uncertainty of $1.6 \times 10^{-6}$. We consider this shift a representative correction for the long-range wavelength distortions of the total spectrum since 84$\%$ of the fiducial $\Delta\mu/\mu$ accuracy is gained from the 2004 exposures.

Creating a 1D spectrum from multiple exposures and overlapping echelle orders involves redispersion onto a single wavelength grid. The rebinning causes correlation in flux and flux uncertainty values between adjacent pixels. Varying the dispersion bin size from the default value of $2.5$ km/s by $\pm0.1$ km/s results in an average $\Delta\mu/\mu$ shift of $\pm 4 \times 10^{-6}$.

The spectra were obtained using 1.0 and 0.8\arcsec~slitwidths in 2004 and 2013, respectively. This corresponds to respective spectral resolutions of $R\sim38\,700$ and $\sim48\,500$ or Gaussian instrumental profiles with the width of $\sigma_{\textrm{inst}}=3.3$ and 2.6~km/s. However, the profile to be convolved with the fitted line profile will be up to $\sim$10\,\% narrower because the quasar light is concentrated towards the centre of the slit. The fiducial $\Delta\mu/\mu$ constraint was derived from the combined 2004 and 2013 spectra using an instrumental profile of $\sigma_{\textrm{inst}}=2.6$ km/s. Varying the $\sigma_{\textrm{inst}}$ in the range from 2.6 to \mbox{2.9 km/s} resulted in deviations from the fiducial $\Delta\mu/\mu$ constraint as large as $3 \times 10^{-6}$.

In summary, the constraint on $\Delta\mu/\mu$ derived from this quasar spectrum shows sensitivity to these systematic effects. All the contributions combined (added in quadrature) constitute the total systematic error budget of $\sigma_{\textrm{sys}} = 5.3\times 10^{-6}$. After correction for the long-range wavelength distortions, our fiducial constraint therefore becomes $\Delta\mu/\mu=(-9.5\pm 5.4_{\textrm{stat}}\pm 5.3_{\textrm{sys}})\times 10^{-6}$.

\begin{figure}
\resizebox{0.5\textwidth}{!}{\includegraphics{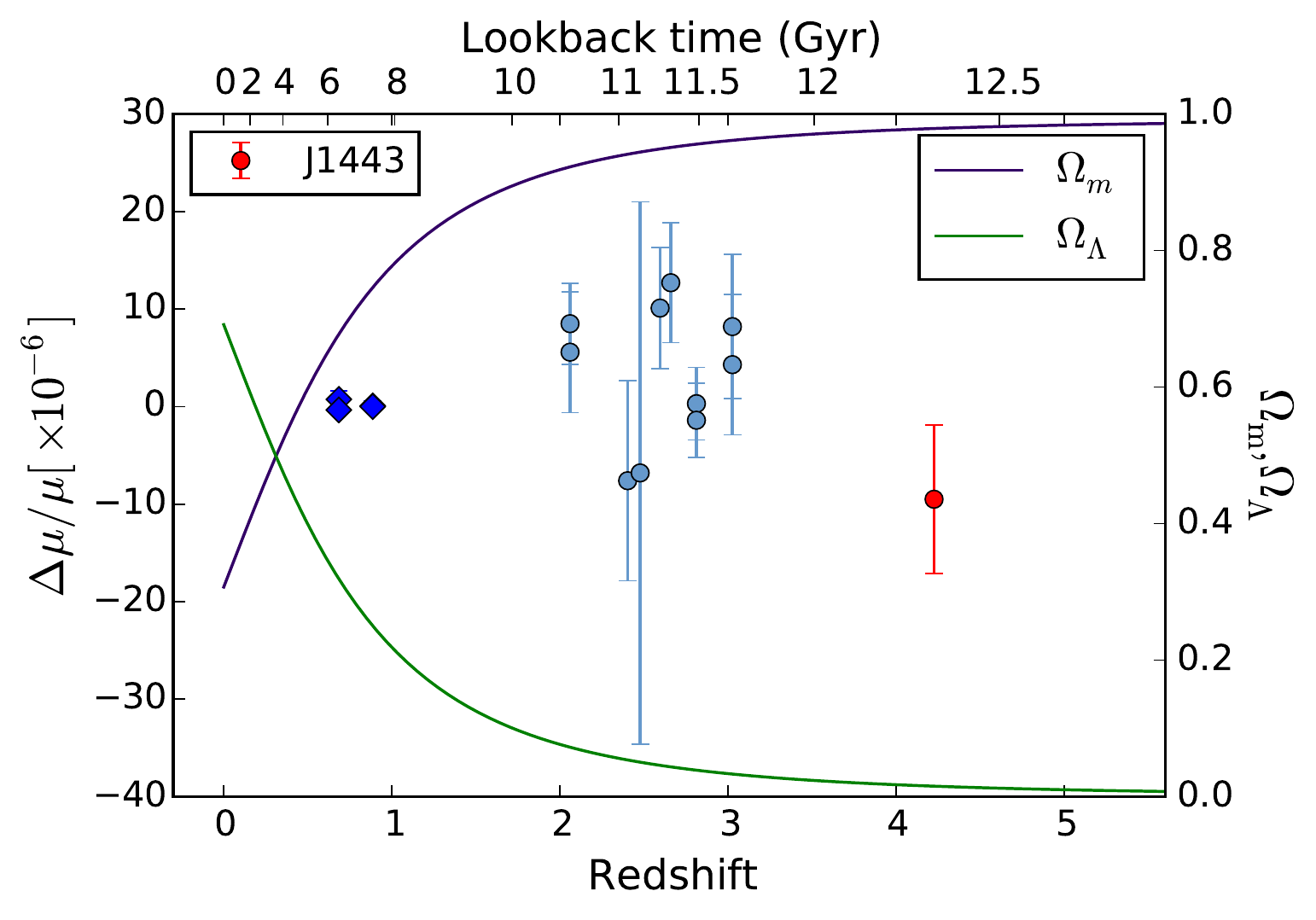}}
\caption{The non-relativistic mass density parameter $\Omega_m$ and the dark energy density parameter $\Omega_{\Lambda}$ as functions of redshift, with a corresponding lookback time axis on top. Results from various $\Delta\mu/\mu$ studies: the diamond markers refer to the measurements of rotational transitions of methanol and ammonia \cite{Murphy2008, Henkel2009, Kanekar2011,Bagdonaite2013}, while $\Delta\mu/\mu$ values derived from H$_2$ observations are denoted by circles \cite{King2008, Malec2010, Bagdonaite2012, Weerdenburg2011, King2011, Wendt2012, Rahmani2013, Bagdonaite2014}.}
\label{fig4}
\end{figure}

This result constrains a possible variation of the proton--electron mass ratio by means  of the H$_2$ spectroscopic method for the highest redshift so far. Fig. \ref{fig4} displays a comparison with previous data, mainly showing results derived from H$_2$ absorbers toward quasars in the redshift interval $z=2-3$, as well as the more constraining results from radio astronomy of polyatomic molecules for $z<0.9$. With the analysis of the absorber toward J1443$+$2724 the window $z>4$ is opened. A further comparison can be made with a result from radio astronomy of a lensed galaxy probing a (7-6) rotational transition in CO and a fine-structure transition in atomic carbon ($^3$P$_2$ - $^3$P$_1$) at $z=5.2$ corresponding to 12.7~Gyr lookback time~\footnote{To compute lookback times we adopt the following values of cosmological parameters: Hubble constant \mbox{H$_{\textrm{0}}=67.3$~km/s/Mpc}, $\Omega_{\textrm{m,0}} = 0.307$, $\Omega_{\Lambda,\textrm{0}} = 0.691$.}; that study probes the combination of dimensionless constants $F=\alpha^2/\mu$ and yields $\Delta F/F< 2 \times 10^{-5}$ \cite{Levshakov2012}.

The presented H$_2$ constraint signifies a null result. It will assist in setting boundaries to various theories describing physics beyond the Standard Model \cite{Bekenstein2002,Sandvik2002}, as well as on the $\Lambda$CDM standard model of cosmology~\cite{Thompson2013}. The densities of matter (including dark matter) $\Omega_{\textrm{m}}$ and dark energy $\Omega_{\Lambda}$, and their ratio (see Fig. \ref{fig4}) are important parameters in the models~\cite{Carroll1998}, and the present work covers a wide set of values: in the interval $z=0 - 4.22$ dark energy covers $\Omega_{\Lambda} = 0.68$  to $0.015$. While a number of cosmological scenarios suggest varying constants \cite{Sandvik2002,Ferreira2014} no quantification of the rate of change is predicted, except for some model-dependent scenarios involving screening and not holding for cosmological time scales~\cite{Brax2014}. Thus, while a clear threshold for new physics is lacking, the observational and experimental targets are positioned to set ever tighter constraints on varying constants. The present study pushes a tight constraint on a varying $\mu$ to lookback times of 10$\%$ of the age of the universe today.

The authors thank the Netherlands Foundation for Fundamental Research of Matter (FOM)
for financial support. M.T.M. and J.B.W. thank the Australian Research Council for Discovery Project grant DP110100866 which supported this work. The work is based on observations with the ESO Very Large Telescope at Paranal (Chile).

\pagebreak
\setcounter{figure}{0}
\setcounter{table}{0}
\setcounter{page}{1}
\makeatletter
\renewcommand{\thetable}{S\arabic{table}}
\renewcommand{\thefigure}{S\arabic{figure}}
\renewcommand{\bibnumfmt}[1]{[S#1]}
\renewcommand{\citenumfont}[1]{S#1}

\input{appendix}

\end{document}

%% file: appendix.tex
\clearpage
\widetext
\begin{center}
\noindent\textbf{\large Supplemental Material: A constraint on a varying proton--electron mass ratio 1.5 billion years after~the~Big~Bang}
\end{center}

\section{Fitted transitions}
\begin{figure*}
\centering
\resizebox{0.9\textwidth}{!}{\includegraphics{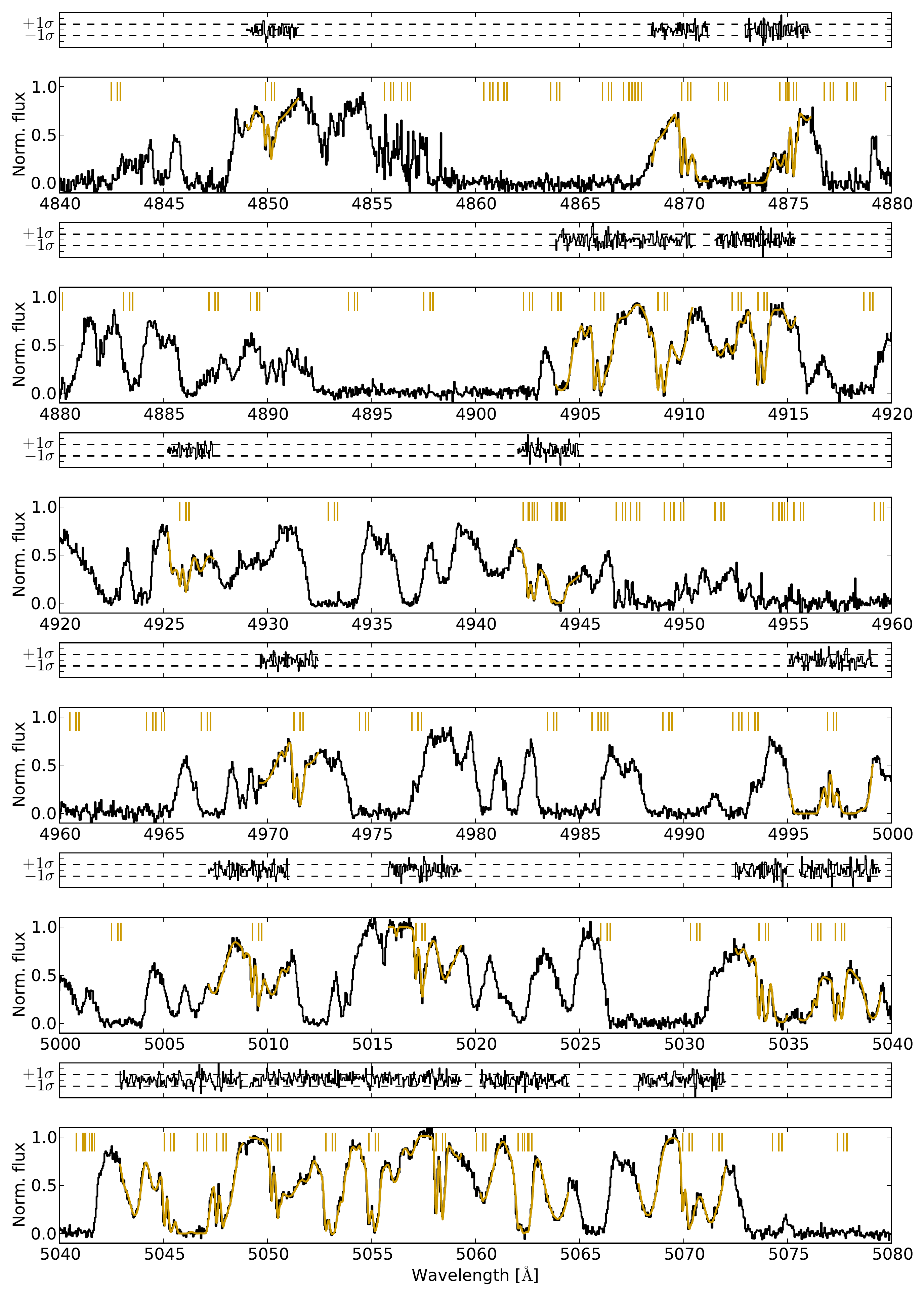}}
\caption{Spectrum of the J1443$+$2724 quasar with the fitted H$_2$ absorption lines at redshift $z=4.224$ (part 1 of 4). The vertical tick marks indicate positions of the velocity components for the $J=(0-4)$ transitions in the 3 VC model.}
\label{appendix:fig2}
\end{figure*}

\begin{figure*}
\centering
\resizebox{0.9\textwidth}{!}{\includegraphics{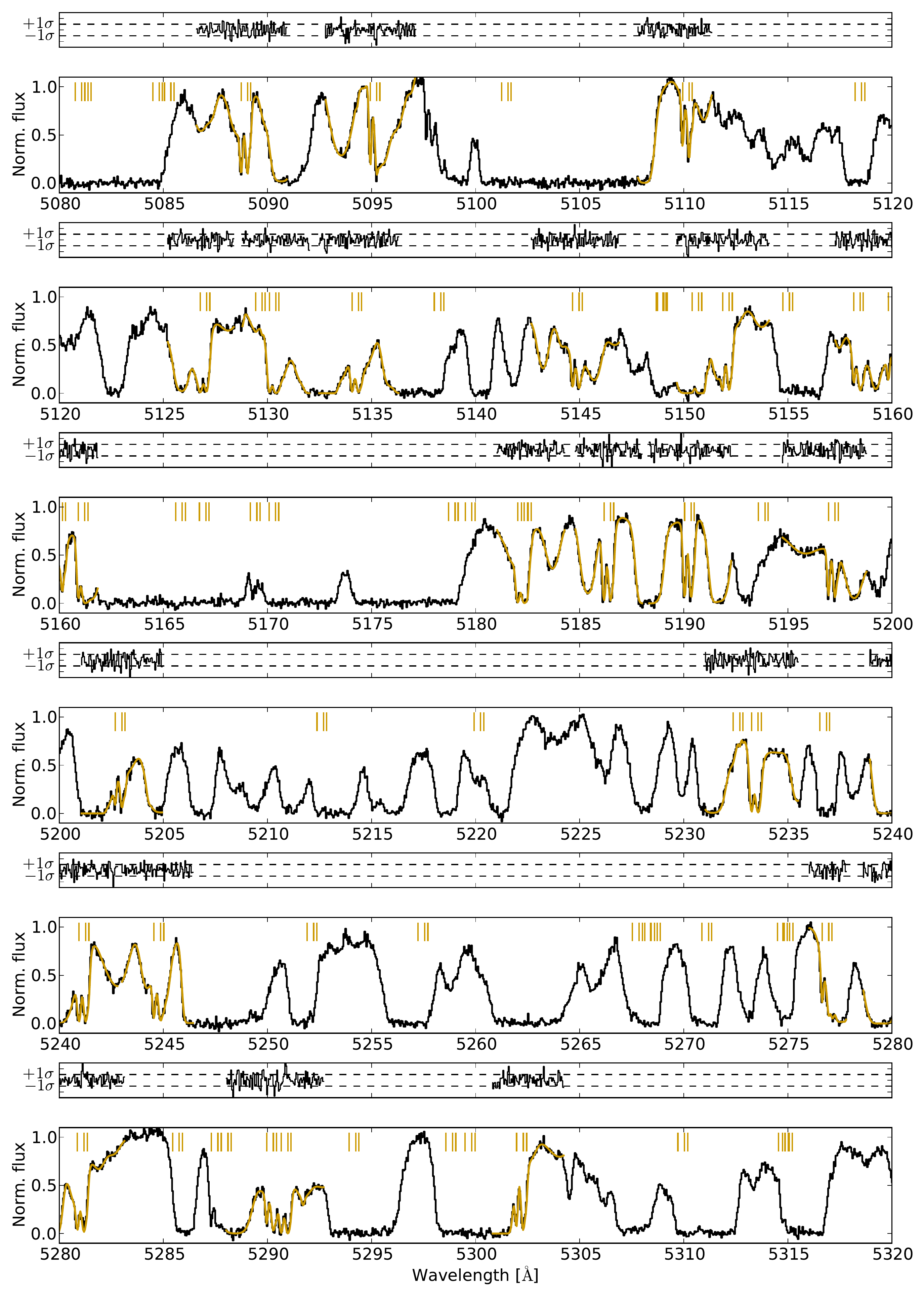}}
\caption{Spectrum of the J1443$+$2724 quasar with the fitted H$_2$ absorption lines at redshift $z=4.224$ (part 2 of 4).}
\label{appendix:fig3}
\end{figure*}

\begin{figure*}
\centering
\resizebox{0.9\textwidth}{!}{\includegraphics{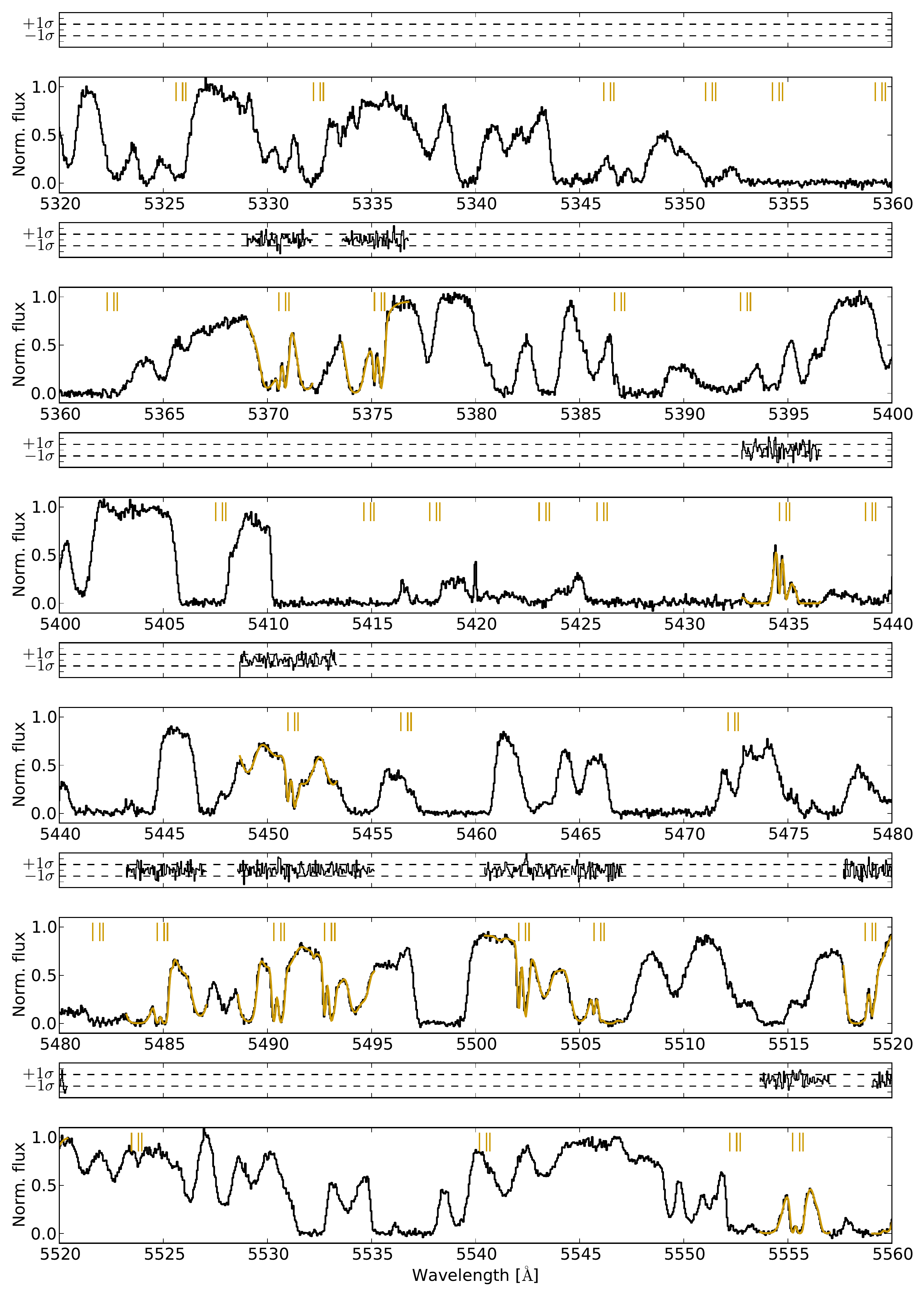}}
\caption{Spectrum of the J1443$+$2724 quasar with the fitted H$_2$ absorption lines at redshift $z=4.224$ (part 3 of 4).}
\label{appendix:fig4}
\end{figure*}

\begin{figure*}
\centering
\resizebox{0.9\textwidth}{!}{\includegraphics{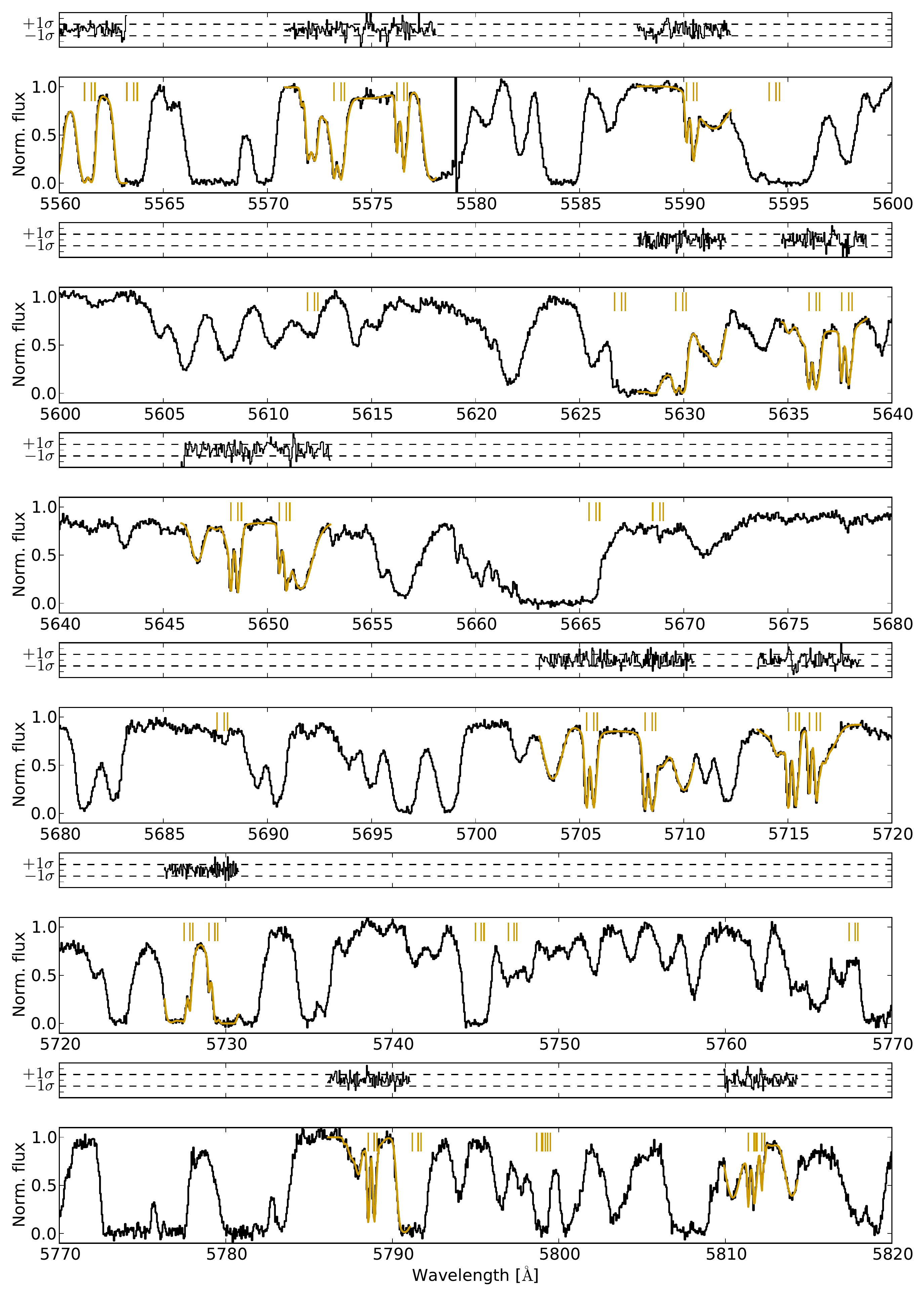}}
\caption{Spectrum of the J1443$+$2724 quasar with the fitted H$_2$ absorption lines at redshift $z=4.224$ (part 4 of 4).}
\label{appendix:fig5}
\end{figure*}

In Table \ref{appendix:table4}, the 89 fitted H$_2$ transitions are listed, grouped by the ground $J$ level. Of these transitions, 17 are from the Werner band. The fitted regions are displayed in Fig. \ref{appendix:fig2} to \ref{appendix:fig5}.
\begin{table}[hb]
	\caption{89 fitted H$_2$ transitions in 60 regions of the J1443$+$2724 spectrum.}
	\label{appendix:table4}
    \begin{tabular}{c@{\hspace{5pt}}c@{\hspace{5pt}}l@{\hspace{1pt}}}

	\toprule
     $J$ level & Number of tr.   & Transition names\\ 
	\colrule
	0 & 7 & W3R(0), L14R(0), L10R(0), L8R(0), \\
	&&L7R(0), L1R(0), L0R(0) \\
	1 & 20 & L16P(1), L15P(1), L15R(1), W3R(1), \\
	&& L12R(1), L12P(1), W2Q(1), L10R(1), \\
	&& L10P(1), L9R(1), L9P(1), L8P(1), \\
	&& L4R(1), L4P(1), L3R(1), L3P(1), \\
	&& L2R(1), L2P(1), L1R(1), L1P(1) \\
	2 & 27 & L16R(2), L15R(2), L13R(2), L12P(2), \\
	&& W2R(2), W2Q(2), W2P(2), L11R(2), \\
	&& L11P(2), L10P(2), W1R(2), W1Q(2), \\
	&& L9R(2), L9P(2), L8R(2), W0Q(2), \\
	&& L7R(2), L6P(2), L5P(2), L4R(2), \\
	&& L4P(2), L3P(2), L2R(2), L2P(2), \\
	&& L1R(2), L1P(2), L0P(2) \\
	3 & 26 & L15R(3), W3P(3), L13R(3), L13P(3), \\
	&& L12R(3), W2R(3), L12P(3), W2Q(3), \\
	&& W2P(3), L11P(3), L10R(3), L10P(3), \\
	&& W1R(3), L9R(3), W0R(3), W0Q(3), \\
	&& L6R(3), L5P(3), L4R(3), L4P(3), \\
	&& L3R(3), L3P(3), L2R(3), L1R(3),\\
	&& L0R(3) \\
	4 & 9 & L17P(4), L16P(4), L15R(4), L13P(4), \\
	&&L12R(4), L11P(4), W1Q(4), L9P(4), \\
	&&W0Q(4) \\

	\botrule
    \end{tabular}
\end{table}

\section{Continuum error}

One possible source of uncertainty in $\Delta\mu/\mu$ can in principle be caused by incorrect treatment of the continuum; we address it here. One can assume that, for 60 fitted sections (containing the 89 H$_2$ lines), the continuum error is random, with the error in the slope of the continuum over any individual H$_2$ section (of width $\sim$100\,km/s) being Gaussian-distributed with a sigma of approximately 0.05 [norm. flux units] / 100 km/s = 5$\times10^{-4}$ per km/s. Individual H$_2$ transitions are $\sim$40\,km/s wide in this absorber, so this corresponds to an incorrect continuum slope of $\sim$2$\times10^{-3}$ across the transition. To a reasonable approximation, this should impart a spurious shift to the fitted centroid of ~0.002 times the width, i.e. 40\,km/s again, or a shift of $\sim80$\,m/s for an individual line. Some of the 89 lines are blended together and therefore share the same spurious shift. Thus, effectively a sample of $\sim$70 transitions reduces the per-line error of 80\,m/s down to a total error budget of 80$/\sqrt{70} \approx 10$\,m/s, which is well below our systematic or random error budgets.

The global quasar continuum assumed initially is just a nominal starting guess which is refined by introducing local continua to each fitting region. These local continua are either constants or straight lines so that degeneracies with broad Ly-$\alpha$ lines would be avoided. Therefore, errors in the global continuum level in each fitting section are marginalized over by fitting a local continuum with a slope and also fitting the broader HI Ly-$\alpha$ lines.

\section{Varying the minimum linewidth parameter}

\begin{figure}
\centering
\resizebox{0.45\textwidth}{!}{\includegraphics{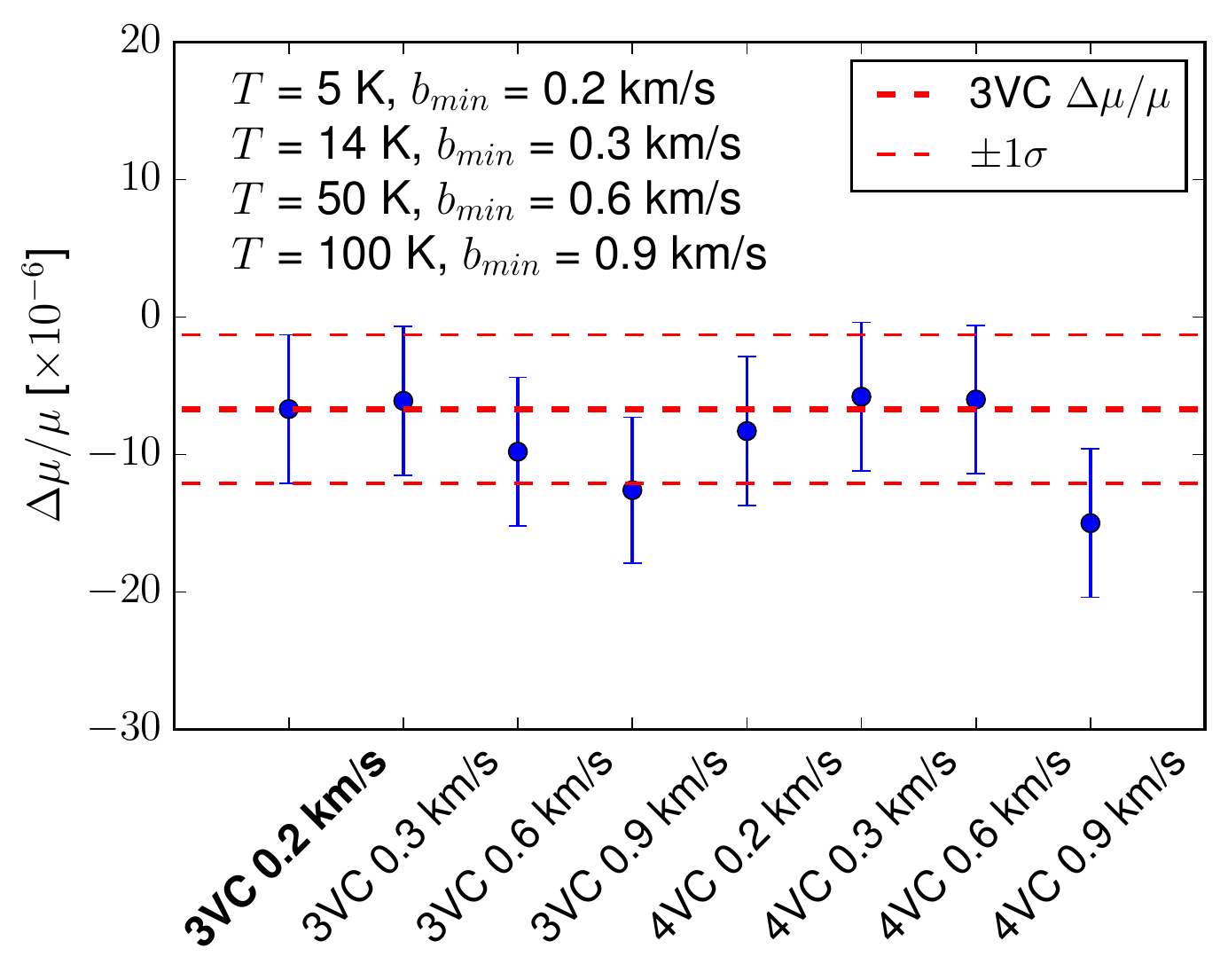}}
\caption{Results of fitting the J1443$+$2724 spectrum with different minimum width parameter $b_{\textrm{min}}$. For the H$_2$ molecule, the Cosmic Microwave Background radiation at \mbox{$z=4.224$} corresponds to $T=14$\,K. The kinetic temperature in cold molecular clouds typically amounts to $50-100$ K.}
\label{appendix:fig-varb}
\end{figure}

The minimum linewidth paramater $b_{\textrm{min}}$ in \textsc{vpfit} is user-defined. Throughout the paper, we use a $b_{\textrm{min}}=0.2$\,km/s. In the case of J1443$+$2724, one of the H$_2$ velocity components reaches this limit. This is likely unphysical because $b_{\textrm{H2}}=0.2$\,km/s corresponds to $T\sim$5\,K, while the temperature of the Cosmic Microwave Background radiation at \mbox{$z=4.224$} corresponds to $T=14$\,K or $b_{\textrm{H2}}=0.3$\,km/s. Furthermore, the kinetic temperature in cold molecular clouds is typically around $\sim100$\,K or $b_{\textrm{H2}}=0.9$\,km/s. In Fig.~\ref{appendix:fig-varb}, we show how imposing these values as $b_{\textrm{min}}$ affects $\Delta\mu/\mu$. Overall, the resulting $\Delta\mu/\mu$ values are in agreement with the fiducial $\Delta\mu/\mu$ limit from the 3VC model with $b_{\textrm{min}}=0.2$\,km/s. However, the goodness of fit parameter $\chi_{\nu}^2$ increases with increasing $b_{\textrm{min}}$ values. For example, a $\chi_{\nu}^2$ of 1.161 results from the 3VC model with $b_{\textrm{min}} = 0.2$\,km/s, and $\chi_{\nu}^2 = 1.189$ from the same model with $b_{\textrm{min}} = 0.9$\,km/s. This indicates that the former model is statistically preferred over the latter.

\begin{figure}
\centering
\resizebox{0.40\textwidth}{!}{\includegraphics{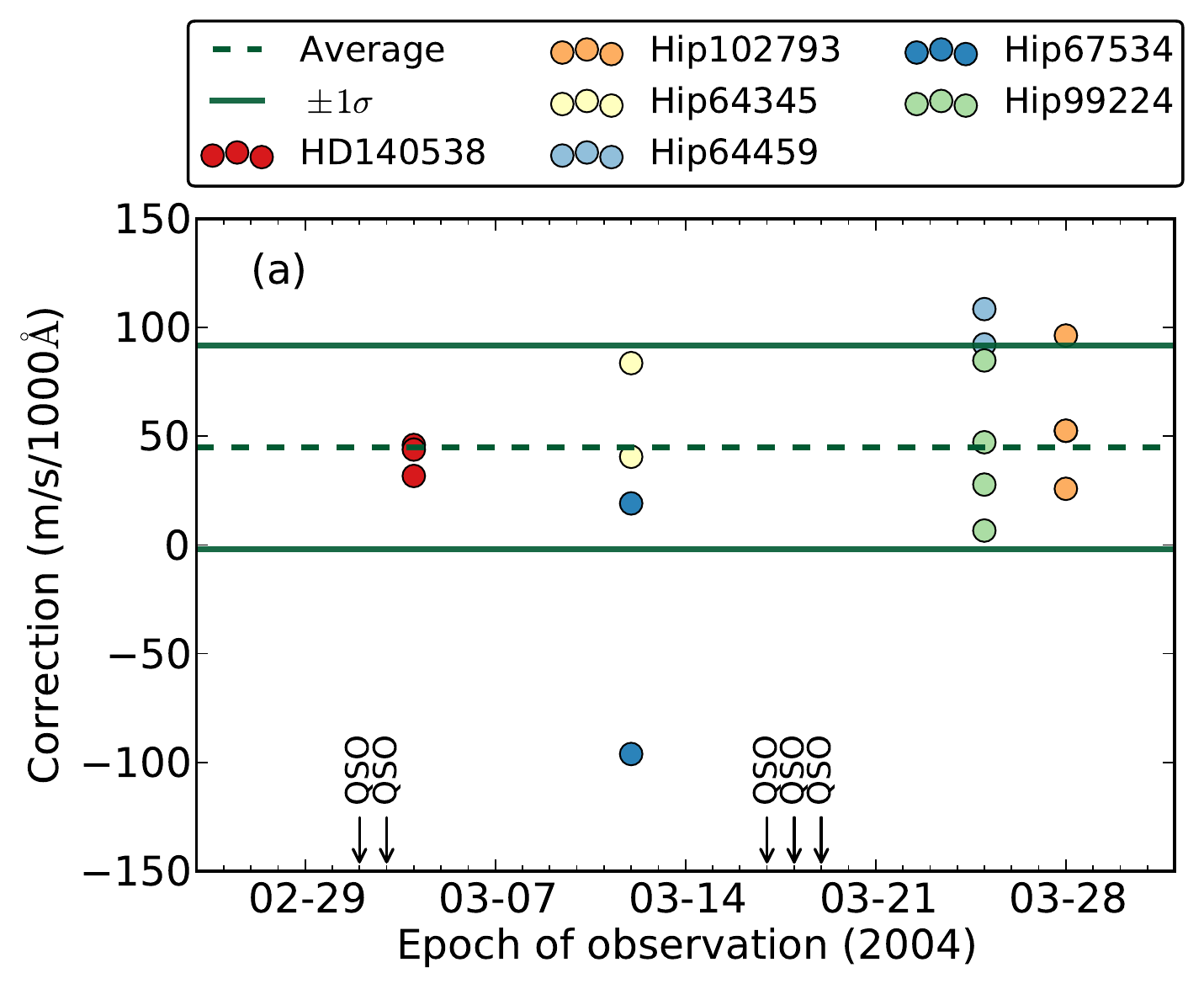}}
\resizebox{0.42\textwidth}{!}{\includegraphics{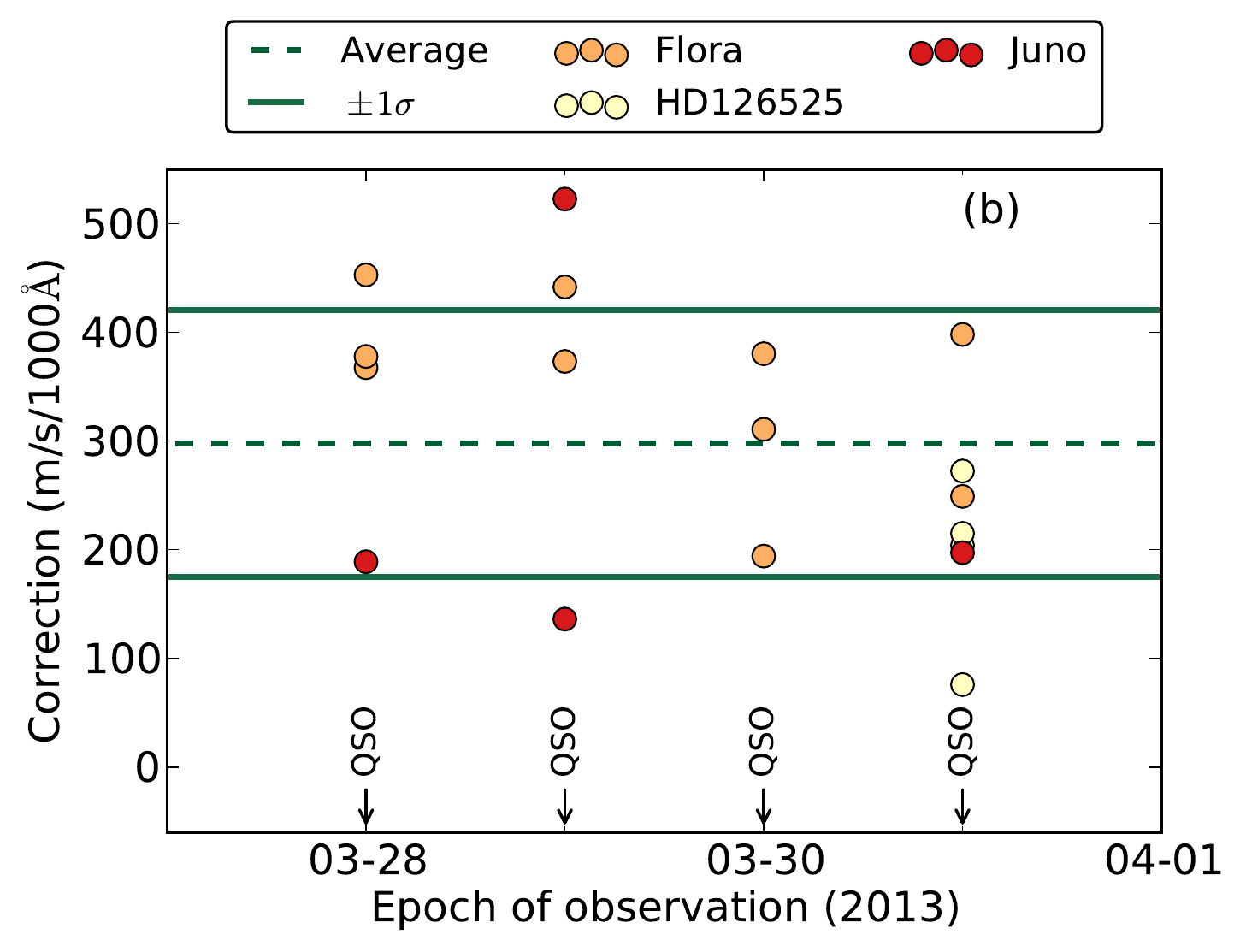}}
\caption{(a) Supercalibration results from 2004. The average correction value amounts to 44.9$\pm$46.8 m/s/1000\AA. (b) Supercalibration results from 2013. The average correction value amounts to 297.6$\pm$122.7 m/s/1000\AA.}
\label{appendix:fig1}
\end{figure}

\section{Supercalibration}

Long-range wavelength distortions in UVES can be quantified by means of a so-called supercalibration method which is based on observations of objects with well understood spectra~[19, 32]. In particular, these objects can be asteroids or solar twins which both exhibit solar-like spectra. Their observations include the usual ThAr calibration routine, used also for the quasar, and the resulting spectra are compared to a highly accurate solar spectrum obtained by an independent instrument, namely Fourier Transform Spectrometer (FTS). The comparison provides means to verify correctness of the ThAr calibration and to make adjustments if it is found to be flawed.

In Table~\ref{appendix:table1}, observational details are provided regarding the J1443$+$2724 observing runs both in 2004 and 2013. The service mode J1443$+$2724 observations were conducted during 2-3 and 17-19 of March, 2004. Note that most of the total SNR of the J1443$+$2724 spectrum originates from this dataset (see Table~\ref{appendix:table3}). Even though dedicated asteroid or solar twin observations were not carried out along with the quasar observations, incidentally, on the ESO archive~[33] we could find several sun-like stars observed during that time period under a different program; their observational details are collected in Table~\ref{appendix:table2}. It has to be noted that even though the same arm of the spectrograph was used, the settings of the quasar observations and those of the sun-like stars were not identical. By applying the supercalibration method described above, we found corrections amounting to 44.9$\pm$46.8 m/s/1000\AA~(see Fig.~\ref{appendix:fig1}). Here we assume that the distortions remain semistable over time periods as long as several weeks, and that the slopes obtained from the solar twins hold also for the main target. The resulting $\Delta\mu/\mu$ correction is quoted in the manuscript. 

Asteroids Juno and Flora, and the solar twin HD126525 were observed along with J1443$+$2724 in March 2013. The instrumental settings of the quasar observations were maintained during the observations of calibration sources: a 544 or 520 nm central wavelength, 2x2 CCD binning, 0.8\arcsec~slit. For this sample of solar-like spectra, a significant deviation from perfect calibration was found with an average slope value of the correction amounting to $297.6\pm 122.7$ m/s/1000\AA. In a study of H$_2$ toward B0642$-$5038, a miscalibration of similar size yielded a $\Delta\mu/\mu$ correction of $-10\times 10^{-6}$~[24]. Fitting the uncorrected 2013 subspectrum results in $\Delta\mu/\mu=(4.0\pm12.4)\times10^{-6}$ for the 3VC model. If a correction is applied a $\Delta\mu/\mu$ of $(-9.3\pm12.4)\times10^{-6}$ is obtained. Because of a low SNR, it is difficult to reliably estimate the uncertainty of this shift. Hence, we do not correct the present 2013 subspectrum of J1443$+$2724.

\begin{table*}
\caption{ESO archival data of the J1443$+$2724 observations with VLT/UVES (program IDs: 072.A-0346(B) and 090.A-0304(A)). The archival exposure name contains the date and time of the observations.  }
\label{appendix:table1}
\begin{tabular}{l@{\hspace{10pt}}c@{\hspace{10pt}}c@{\hspace{10pt}}c@{\hspace{10pt}}c@{\hspace{4pt}}}
\toprule
Archival exposure name & Central wavelength & CCD binning & Slitwidth & Integ. time\\
  & [nm] &  & [\arcsec] & [s]\\
\colrule

UVES.2004-03-02T07:52:43.078.fits & 580  & 2x2 &  1.0 &  5225.00  \\
UVES.2004-03-03T08:12:39.143.fits & 580  & 2x2 &  1.0 &  5016.00\\
UVES.2004-03-17T08:14:13.810.fits & 580  & 2x2 &  1.0 &  5091.00 \\
UVES.2004-03-18T06:53:02.273.fits & 580  & 2x2 &  1.0 &  5225.00 \\
UVES.2004-03-19T06:35:36.823.fits & 580  & 2x2 &  1.0 &  5800.00 \\
UVES.2013-03-28T05:15:58.805.fits & 544  & 2x2 &  0.8 & 4800.00  \\
UVES.2013-03-28T06:39:39.743.fits & 544  & 2x2 &  0.8 & 4800.00  \\
UVES.2013-03-28T08:02:50.876.fits & 544  & 2x2 &  0.8 & 4800.00  \\
UVES.2013-03-29T05:22:20.681.fits & 544  & 2x2 &  0.8 & 4800.00  \\
UVES.2013-03-29T06:45:12.579.fits & 544  & 2x2 &  0.8 & 4800.00  \\
UVES.2013-03-29T08:10:27.130.fits & 544  & 2x2 &  0.8 & 4800.00  \\
UVES.2013-03-30T05:03:29.905.fits & 520  & 2x2 &  0.8 & 4800.00 \\
UVES.2013-03-30T06:28:32.340.fits & 520  & 2x2 &  0.8 & 4800.00 \\
UVES.2013-03-30T07:51:11.513.fits & 520  & 2x2 &  0.8 & 6000.00 \\
UVES.2013-03-31T05:03:05.640.fits & 544  & 2x2 &  0.8 & 4800.00 \\
UVES.2013-03-31T06:25:58.450.fits & 544  & 2x2 &  0.8 & 4800.00 \\
UVES.2013-03-31T07:49:11.563.fits & 544  & 2x2 &  0.8 & 6300.00 \\
\botrule
\end{tabular}
\end{table*}

\begin{table*}
\centering
\caption{Representative signal-to-noise (SNR) values measured in the continuum of the J1443$+$2724 spectrum. The redshifted H$_2$ transitions are found in the range between 484 and 581 nm. The total spectrum is a combination of two datasets obtained independently in 2004 and 2013. The SNR at shorter wavelengths is entirely dominated by the data from 2004 as the observations in 2013 were affected by the full Moon.}
\label{appendix:table3}
\begin{tabular}{c@{\hspace{40pt}}c@{\hspace{40pt}}c@{\hspace{20pt}}}
\toprule
 $\lambda$ [nm] & SNR$_{tot}$ & SNR$_{2004}$ \\
\colrule
    501.5           & 25               & 23                \\
    528.5           & 30               & 26                \\
    560.0           & 48               & 39                \\
\botrule
\end{tabular}
\end{table*}

\begin{table*}
\caption{ESO archival data of several sun-like stars observed in 2004 (program IDs: 072.D-0739(A) and 072.B-0179(A)), at a similar time when also part of the J1443$+$2724 data were taken. The archival exposure name contains the date and time of the observations.}
\label{appendix:table2}
\begin{tabular}{l@{\hspace{10pt}}l@{\hspace{10pt}}c@{\hspace{10pt}}c@{\hspace{10pt}}c@{\hspace{4pt}}}
\toprule
Target & Archival exposure name & Central wavelength & CCD binning & Slitwidth \\
 &  & [nm] &  & [\arcsec] \\
\colrule
HD140538 & UVES.2004-03-04T08:11:03.429.fits & 860  & 2x2  & 1.0 \\
HD140538 & UVES.2004-03-04T08:11:06.685.fits & 390  & 2x2  & 1.0  \\
Hip64345 & UVES.2004-03-12T07:14:53.472.fits & 564 &  1x1 &  0.3  \\
Hip67534  & UVES.2004-03-12T08:09:56.593.fits & 564 & 1x1 & 0.3  \\
Hip64459 & UVES.2004-03-25T07:43:09.439.fits & 564 & 1x1 & 0.3  \\
Hip64459 & UVES.2004-03-25T07:46:30.774.fits & 564 & 1x1 &  0.3  \\
Hip99224 & UVES.2004-03-25T09:12:28.662.fits & 564 & 1x1 & 0.3  \\
Hip99224 & UVES.2004-03-25T09:22:46.173.fits  & 564  & 1x1 & 0.3  \\
Hip102793 & UVES.2004-03-28T09:01:15.687.fits  & 564 & 1x1 & 0.3  \\
Hip102793 & UVES.2004-03-28T09:17:43.306.fits  & 564 & 1x1 & 0.3  \\
\botrule
\end{tabular}
\end{table*}
